\documentclass{ccjnl}
\usepackage{subcaption}

\title{C-RAN Advanced: From a Network Cooperation Perspective}
\author{Xiaoyun Wang, Yutong Zhang\inst{*}, Sen Wang, Qi Sun, Hanning Wang, Qixing Wang, Jing Jin, Jiwei He, Nan Li\corinfo{zhangyutongyj@chinamobile.com}}
\receiveddate{April 21, 2025}
\reviseddate{August 12, 2025}
\Editor{****}

\address[]{China Mobile Research Institute, Beijing 100053, China}

\begin{document}
\maketitle

\begin{abstract}
Future mobile networks in the sixth generation (6G) are poised for a paradigm shift from conventional communication services toward comprehensive information services, driving the evolution of radio access network~(RAN) architectures toward enhanced cooperation, intelligence, and service orientation.
Building upon the concept of centralized, collaborative, cloud, and clean RAN (C-RAN), this article proposes a novel cooperative, intelligent, and service-based RAN (CIS-RAN) architecture.
Focusing on cooperation, CIS-RAN extends the traditional cooperative communication paradigm by further integrating cooperative sensing and cooperative artificial intelligence (AI).
To improve both performance and effectiveness across diverse application scenarios, CIS-RAN enhances network cooperation throughout the entire process of acquisition, transmission, and processing, thereby enabling efficient information acquisition, diverse cooperative interactions, and intelligent fusion decision-making.
Key technologies are discussed, with network cooperative multiple-input multiple-output~(MIMO) examined as a case study, demonstrating superior performance over traditional architectures, as demonstrated by numerical results.
Future research directions are outlined, emphasizing the continued exploration and advancement of the CIS-RAN architecture, particularly in enhancing network cooperation.
\end{abstract}

\section{introduction}\label{sec:intro}
\vspace{-1em}
Driven by the rapid surge in mobile traffic, China Mobile introduced the concept of centralized, collaborative, cloud, and clean radio access network (C-RAN) in 2010~\cite{wang2010c,chen2011c}, representing a significant advancement in radio access network~(RAN) architecture and providing a cost-efficient solution to meet the growing demands for higher data transmission rates and improved service quality~\cite{xiaoyun2010c,huang2015white}.
Specifically, the C-RAN architecture aggregates all baseband units (BBUs) from distributed base stations (BSs) into a central pool, enabling the shift from single-point optimization to \textit{multi-point cooperation}~\cite{chen2013c}.
Leveraging cooperative network architecture and virtualization technologies, such a centralized deployment reduces site rental costs, capital expenditures, and operational expenses~\cite{LarsenSurvey}.
As a result, it enables rapid deployment and scalability, effectively addressing the growing challenges of wireless communication networks~\cite{checko2014cloud}, including reduced energy consumption, lower operational and maintenance costs, improved spectral efficiency, and ultimately supporting sustainable development~\cite{wu2015cloud}.

As the deployment and commercial operation of C-RAN architecture are speeding up, novel RAN solutions have been exploited with respect to flexible interfaces and on-demand decision-making in future networks to achieve network cooperation across broader geographical range.
To address the limitations in bandwidth and architectural flexibility, China Mobile proposed the concept of a next-generation fronthaul interface~(NGFI), aiming for a packet-based, traffic-dependent, and antenna scale-independent interface~\cite{chih2015rethink}.
The key to realizing NGFI lies in the splitting of baseband functions, explored through the division of the BBU into the central unit (CU) and the distributed unit (DU)~\cite{chih2018ran}, ensuring compatibility of RAN.
To enable on-demand decision-making, data processing in existing RAN is distributed across different nodes according to various requirements.

However, future mobile networks in the sixth generation (6G) are poised for a revolutionary shift from traditional communication services to comprehensive information services, posing new challenges in performance enhancement, capability expansion, and flexible service provisioning~\cite{jiang2021road}.
As outlined in the ITU-R recommendations, 6G will foster a broader range of usage scenarios, including integrated artificial intelligence~(AI) and communications, integrated sensing and communication~(ISAC), as well as higher key performance indicators~(KPIs), such as spectral efficiency, latency, and reliability~\cite{recommendation2023framework}.
Specifically, in cooperative sensing, multiple network nodes jointly process received signals to improve sensing accuracy, coverage, and robustness~\cite{10557715,10601686,10872780}.
In cooperative AI, 3GPP RAN introduced AI/ML support since Rel.~18 for several RAN optimization use cases, where the collaboration among gNBs is limited to sharing performance metrics~\cite{3gpp2023xn}.
The study is also extended to AI/ML applications on the air interface~\cite{study2023technical}, where further collaboration possibilities are investigated.
Furthermore, the diversified scenarios of 6G highlight the growing demand for customized and differentiated network services, necessitating a paradigm shift from fixed provisioning to flexible service delivery~\cite{schmidt2021ran}.
To meet these emerging requirements, traditional RAN architectures are expected to evolve toward enhanced cooperation, intelligence, and service orientation.

In this article, we introduce the cooperative, intelligent, and service-based RAN (CIS-RAN) architecture by enhancing cooperation, integrating intelligence, and adopting a service-oriented approach.
Focusing on cooperation enhancement, beyond the improvements in network cooperation for communication, existing studies have also investigated cooperative sensing and cooperative AI within specific application scenarios~\cite{dai2022survey, ma2019sensing,nemati2022toward,zhu2023pushing,li2025rethinking,huang2024communication,wang2024integration}, while others have extended network cooperation to either domain in more general settings~\cite{wen2023task,mao2017a,zhu2019mimo,liu2020joint,baltruvsaitis2018multimodal,feng2021joint,cheng2023intelligent,cui2021integrating,liu2022integrated}.
In~\cite{dai2022survey}, the integrated sensing, communication, and AI networks for smart oceans have been discussed with respect to the key technologies, challenges, and applications in maritime networks.
The authors in~\cite{wang2024integration} have investigated the integration of communication, sensing, and AI to address the challenges posed by the conflict between limited resources and user requirements in metaverse scenarios.
The authors in~\cite{mao2017a} have discussed enhancing system performance, such as energy efficiency and latency, by balancing the communication and AI loads on the devices.
The authors in~\cite{liu2020joint} have reviewed the advancements in radar-communication coexistence and dual-functional radar-communication systems, proposing a novel transceiver architecture, frame structure, and a new scheme for joint target search and communication channel estimation.
In~\cite{liu2022integrated}, a comprehensive review has been provided with respect to the background, range of key applications, and state-of-the-art approaches of integrated sensing and communications.

\begin{figure*}[t]
\centering
\subfloat[Single-node MIMO.]
{
\label{Fig:singleMIMO}
\includegraphics[width=0.3\textwidth]{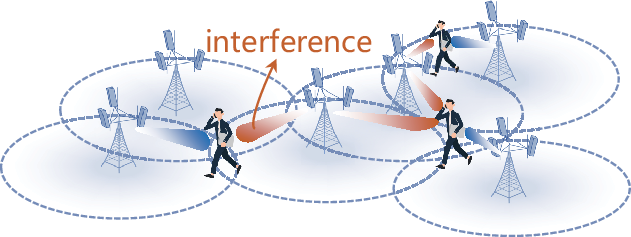}
}
\subfloat[Cooperative MIMO\\(BS-centric).]
{
\label{Fig:networkcentric}
\includegraphics[width=0.3\textwidth]{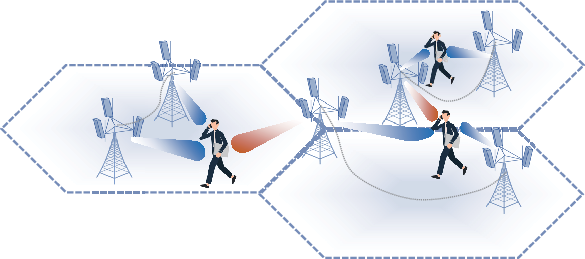}
}
\subfloat[Network cooperative MIMO\\(User-centric).]
{
\label{Fig:usercentric}
\includegraphics[width=0.3\textwidth]{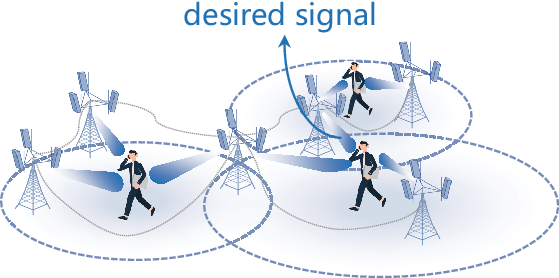}
}
\caption{Evolution towards enhanced network cooperation in communication.}
\label{Fig:intro}
\end{figure*}
Nevertheless, the development of a general RAN architecture for next-generation wireless networks that facilitates network cooperation across communication~\cite{han2023network}, sensing, and AI domains remains under explored.
Therefore, the proposed CIS-RAN architecture enhances the existing cooperative communication while also incorporating cooperative sensing and cooperative AI.
The CIS-RAN architecture is evolved from the following aspects.
\begin{itemize}
  \item \textit{Support user-centric cooperative communication}:
  As shown in Figure~\ref{Fig:intro}, to tackle inter-cell interference, conventional C-RAN typically employ BS-centric cooperative clusters to enhance data rates within predefined coordination regions.
  However, the extremely-high coordination overhead and inherent cluster boundaries severely limit the practical performance gains.
  The proposed CIS-RAN architecture overcomes these limitations by enabling \textit{user-centric} cooperative MIMO.
  Specifically, it constructs \textit{always-on} cooperative clusters that support seamless cooperative transmission and handover, thereby enhancing user-perceived service continuity and driving global network optimality.
  \item \textit{Support extended cooperative sensing/AI}:
  Building upon cooperative communication in C-RAN, the CIS-RAN architecture further extends cooperation to sensing and intelligence.
  Specifically, due to limitations in signal-to-noise ratio, time-frequency resources, and array aperture, single-node sensing fails to achieve the accuracy required by 6G in terms of range, velocity, and angle estimation.
  Enhancing sensing performance across these dimensions necessitates cooperative processing among multiple nodes.
  Similarly, to address the limitations of computing power and data availability, CIS-RAN enables cooperative AI, where computational and data resources are efficiently shared across nodes, not only improving inference accuracy and reducing generalization error, but also enhancing the ubiquity and scalability of AI capabilities.
\end{itemize}

Aiming to realize the vision outlined above, CIS-RAN enhances network cooperation across the entire process including acquisition, transmission, and processing, thereby improving both performance and effectiveness across diverse application scenarios.
The primary contributions of this proposed CIS-RAN architecture are outlined below.

\begin{enumerate}
    \item The CIS-RAN architecture is proposed to enhance network cooperation to support user-centric cooperative communication and extended cooperative sensing/AI.
    
    \item To enhance network cooperation, CIS-RAN proposes several key technologies focused on the entire process of acquisition, transmission, and processing, enabling efficient information acquisition, diverse information cooperation and interactions, and intelligent fusion decision-making, specifically.

    \item Based on the CIS-RAN architecture and key technologies, network cooperative MIMO is studied as an example in terms of channel state information~(CSI) acquisition, dynamic clustering, resource allocation, and precoding design.
        Simulation results show that the proposed approach achieves higher spectral efficiency compared to traditional schemes.

    \item Several opening issues are highlighted that require further exploration of the CIS-RAN architecture, including candidate RAN functional split options and interaction mechanism design.
\end{enumerate}

The rest of paper is organized as follows.
Section~\ref{sec:fundamentals} introduces the fundamentals of CIS-RAN, covering its design requirements, design challenges, network architecture, and technical advantages.
Section~\ref{sec:technologies} presents the key technologies associated with CIS-RAN.
Building upon this, Section~\ref{sec:case} studies network cooperative MIMO as an example, focusing on CSI acquisition, dynamic clustering, resource allocation, and precoding design, and provides simulation results.
In Section~\ref{sec:future}, several opening issues are highlighted that require further exploration of the CIS-RAN architecture.
Finally, we draw our conclusion in Section~\ref{sec:conclusion}.
\vspace{-1.0em}
\section{CIS-RAN Architecture}\label{sec:fundamentals}
In this section, we first discuss the design requirements.
On this basis, the CIS-RAN architecture is proposed to meet the high-performance demands of 6G networks across diverse usage scenarios.
Finally, we introduce the technical advantages of the proposed architecture.
\vspace{-0.5em}
\subsection{Design Requirements}\label{subsec:challenges}

Aiming to meet the evolving and diverse demands of future networks, CIS-RAN is expected to enhance network cooperation through the improvement of cooperative communication, along with the introduction of cooperative sensing and cooperative AI.
In the rest of this subsection, we discuss each of these aspects in detail and analyzes the corresponding architectural requirements driven by such evolutions.
\vspace{-0.5em}
\subsubsection{Support user-centric cooperative communication}
\vspace{-0.5em}
As mentioned in Section~\ref{sec:intro}, the proposed CIS-RAN architecture enables user-centric cooperative MIMO, improving user-perceived service continuity and achieving seamless network performance through the following three key transformations.
\begin{itemize}
  \item \textit{From single-point access to always-on cooperative access}: 
  The system constructs dynamic, user-centric cooperative clusters that remain persistently active, allowing users to access the network instantly and benefit from seamless reconfiguration without service interruption.
  \item \textit{From non-coherent to coherent joint transmission}:
  By improving synchronization accuracy across base stations, the system shifts from non-coherent to coherent joint transmission, significantly enhancing cooperative transmission gains and spectral efficiency.
  \item \textit{From bounded to boundary-Free cooperation}:
  The architecture removes the physical and logical constraints of conventional C-RAN by enabling cooperation across multiple BBUs, thereby supporting uninterrupted service continuity and smooth mobility experiences.
\end{itemize}

\vspace{-0.5em}
\subsubsection{Support extended cooperative sensing/AI}
\vspace{-0.5em}
To support the diverse usage scenarios of 6G, CIS-RAN is expected for further extensive cooperation to sensing and intelligence, thereby enabling the network to dynamically adapt to varying service requirements and achieve enhanced performance across a wide range of applications, such as ISAC and integrated AI and communications, and so on.
Specifically, By leveraging multiple sensing nodes, cooperative sensing jointly process sensing data collected from distributed BSs, which share transmission and reception signals, thereby realizing coordinated sensing signal processing across the network.
To support real-time AI inference and strengthen network cooperation, RAN architectures should incorporate on-demand computing and data processing capabilities.
By enabling cooperative AI, computing and data resources can be shared across nodes, requiring centralized cooperative units for unified data, model, and computation coordination.
The situation slows the standardization work which would enable the exchange of AI/ML models in multivendor deployment.

Given the above demands of future networks, realizing the full potential of such cooperation in large-scale deployments presents several critical challenges.
First, high communication overhead poses a major bottleneck, as sharing raw sensing data or high-dimensional AI features incurs significant fronthaul and backhaul bandwidth consumption.
Second, strict time and frequency synchronization is essential for coherent joint processing, yet maintaining such synchronization across geographically distributed nodes is both technically demanding and cost-intensive.
Finally, data and resource heterogeneity, including disparities in computational capabilities among nodes, introduces further difficulty in efficient data fusion and model aggregation.

Therefore, in CIS-RAN, it is essential to improve synchronization accuracy across multiple nodes.
Building on this, it is critical to enable real-time information exchange among BBUs to support dynamic coordination.
Additionally, the architecture should integrate advanced data acquisition and sensing capabilities, as well as the deployment of multi-point fusion processing units to enable intelligent and efficient cooperation across the network.
\begin{figure*}[t]
	\centering
    \includegraphics[width=0.9\textwidth]{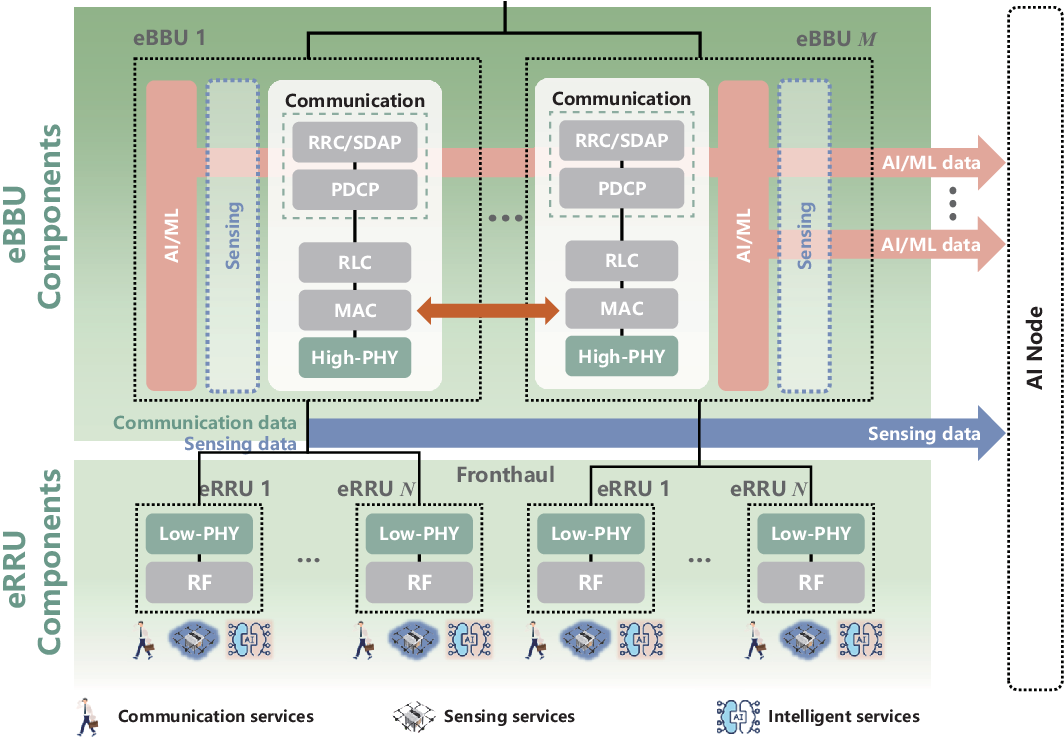}
	\caption{CIS-RAN architecture.}
	\label{Fig:CRANa}
\end{figure*}
\vspace{-1.0em}
\subsection{Architecture Design}
To tackle with the above mentioned issues, as shown in Figure~\ref{Fig:CRANa}, we propose and design the CIS-RAN architecture consisting of the enhanced baseband unit~(eBBU) components\endnote{The eBBU could be further spitted into enhanced CU and enhanced DU, as indicated by the green dashed box in Figure~\ref{Fig:CRANa}}, the enhanced remote radio unit~(eRRU) components, and the AI node where each component is enhanced by integrating of the communication, sensing, and AI capabilities.
The eBBU and eRRU are connected via the fronthaul interface, used for transmitting in-phase and quadrature~(IQ) data and lower-layer control signals.
\begin{itemize}
    \item \textbf{AI node}: To support AI and facilitate multi-dimensional data processing, CIS-RAN establishes an independent AI node that encompasses modules for computing resource management, model lifecycle management (LCM), data collection, and other essential functions.
        On one hand, the AI node processes data from various AI/machine learning~(ML) modules through its interface with eBBUs.
        On the other hand, it is directly connected to the fronthaul, allowing for efficient processing of raw sensing data.
    
    \item \textbf{eBBU}: The eBBU is responsible for centralized control and high-level protocol processing.
    The communication protocol stack includes radio resource control (RRC), service data adaptation protocol (SDBS), packet data convergence protocol (PDCP) radio link control (RLC), medium access control (MAC), and high physical (High-PHY) layer functions, which are responsible for session management, quality of service~(QoS) assurance, data encryption, header compression, redundancy control, egmentation/reassembly and automatic repeat request~(ARQ) retransmission, scheduling resource allocation and hybrid ARQ~(HARQ) management, and high-level physical layer functions such as channel coding and modulation, respectively.
    To support sensing applications, each eBBU may integrate a sensing function module for data fusion processing, where the dashed box indicates one of the options for deploying the sensing function.
    In addition, each eBBU is equipped with AI capabilities, facilitating the distributed processing of both communication and sensing data.
    Furthermore, besides the local intelligent computing power, the eBBU can interact with the AI node for collaborative AI/ML model learning and inference for improved AI/ML performance and computing resource efficiency.
    Besides, direct interactions among BBUs are enabled to support real-time cooperation across BBU boundaries.
    
    \item \textbf{eRRU}: The eRRU layer is responsible for radio frequency (RF) signal processing and wireless signal transmission/reception.
    Specifically, the low physical (Low-PHY) layer includes functions such as fast Fourier transform~(FFT)/inverse fast Fourier transform~(IFFT), beamforming, and data sensing.
    The RF front-end primarily handles analog-to-digital conversion (ADC) and digital-to-analog conversion (DAC) for wireless signal transmission.
    In addition to supporting communication data transmission, the eRRU also facilitates the collection of sensing data.
\end{itemize}

\begin{figure*}[t]
	\centering
    \includegraphics[width=0.8\textwidth]{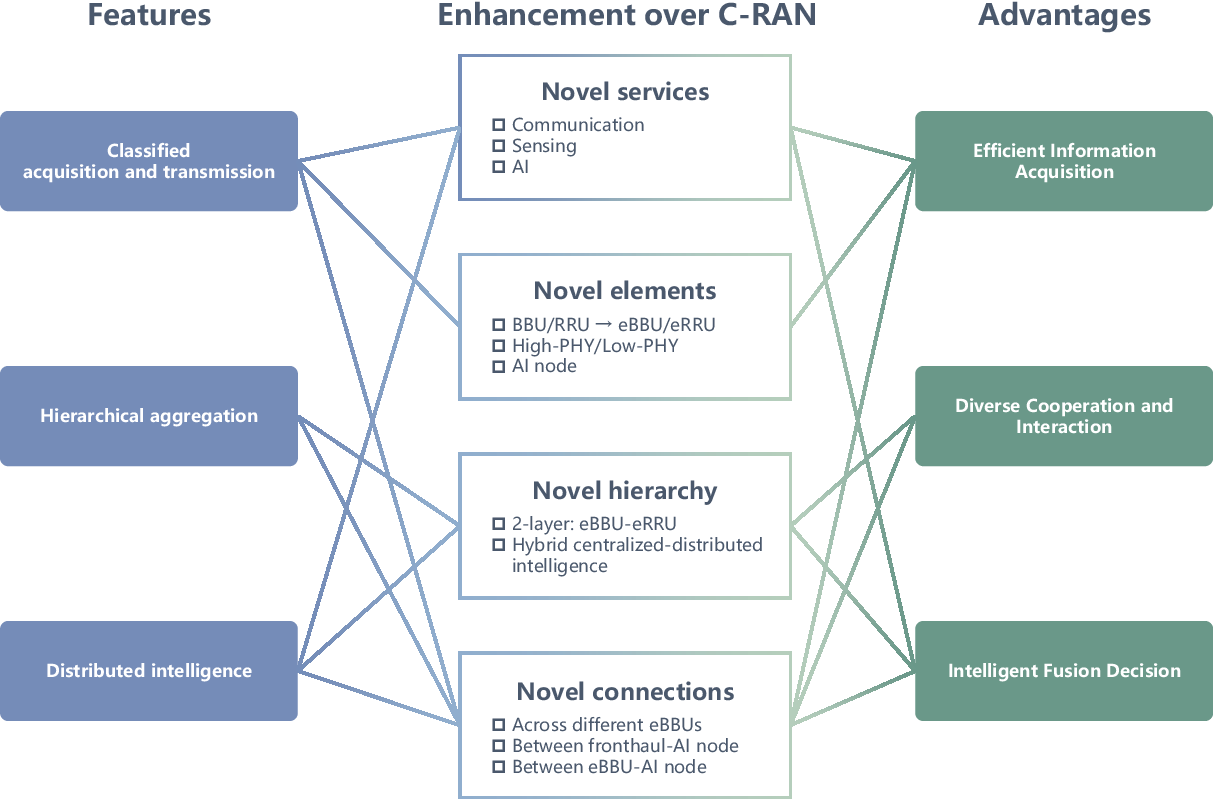}
	\caption{Features, enhancement, and advantages of CIS-RAN architecture over C-RAN.}
	\label{Fig:logic}
\end{figure*}
As shown in Figure~\ref{Fig:logic}, compared to conventional C-RAN architecture, CIS-RAN exhibits novel services, elements, hierarchy, and connections.
Benefiting from the above enhancements, three key features are developed around the full process of network cooperation, including acquisition, transmission, and processing, ultimately enabling superior performance across diverse usage scenarios.
\begin{itemize}
    \item \textit{Classified acquisition and transmission}:  To support various applications, which inherently brings diverse data requirements, the proposed CIS-RAN architecture facilitates classified data acquisition and transmission.
    As shown in the Figure~\ref{Fig:CRANa}, by categorizing data based on its type and processing demands, CIS-RAN enables the parallel acquisition of communication, sensing, and AI data, which are then transmitted through multiple layers.
    \vspace{-0.4em}
    \item \textit{Hierarchical aggregation}: To enable enhanced network cooperation, improved information interaction is crucial, requiring a hierarchical architecture with flexible interfaces that strategically distribute various functions across multiple network layers.
        As shown in the Figure~\ref{Fig:CRANa}, the functionality of high-PHY and low-PHY are redefined where precoding and data sensing are handled at low-PHY, with tasks such as coding carried out at high-PHY.
        Furthermore, eBBUs are allowed to directly exchange information to facilitate real-time cooperation beyond individual boundaries.
        By leveraging such a scalable infrastructure, CIS-RAN also simplifies maintenance and optimizes network resource management, thereby reducing deployment and operational expenses.
    \item \textit{Distributed intelligence}: Based on a hierarchical architecture, CIS-RAN adopts a hybrid centralized–distributed strategy for resource deployment and management, enabling intelligent decision-making.
        Under centralized coordination, computational resources are flexibly allocated across nodes and layers to improve efficiency.
        Specifically, inference tasks are executed at eBBU-side AI/ML modules, while model training is carried out at the AI node.
\end{itemize}
\vspace{-1.0em}
\subsection{Technical Advantages}
\vspace{-0.5em}
As shown in Figure~\ref{Fig:logic}, CIS-RAN leverages classified acquisition and transmission, hierarchical aggregation, and distributed intelligence to enhance the full process of network cooperation, including acquisition, transmission, and processing, thereby delivering the following advantages.
\vspace{-1.0em}
\subsubsection{Efficient Information Acquisition}
\vspace{-0.5em}
Different types of data are acquired and transmitted using tailored strategies, each guided by its own service-level expectations.
By applying priority-based scheduling and allocating resources accordingly, the system avoids unnecessary overhead and reduces interference across data types.
Such a dynamic, demand-driven approach enables cost-effective, accurate, real-time, and comprehensive data acquisition and transmission.
\vspace{-1.0em}
\subsubsection{Diverse Cooperation and Interaction}
\vspace{-0.5em}
Benefiting from the hierarchical architecture and diversified interaction interfaces, CIS-RAN achieves efficient heterogeneous control and transmission under latency, bandwidth, and topology constraints, thereby enhancing network cooperation.
The novel interfaces between eBBUs support fast handover signaling interactions, reducing the handover latency and enhancing user experience.
Specifically, compared with conventional RAN architectures, CIS-RAN with inter-eBBU interfaces achieves a 37\%/67\%/50\%/50\% reduction in the number of messages for the access/handover/dedicated bearer establishment/inactive handover, respectively~\cite{patent1,patent2,patent3}.
Moreover, by decoupling high-PHY and low-PHY for asynchronous task processing, fronthaul bandwidth requirements and transmission latency are further reduced.
The flexible and diverse cooperation and interaction improve the network flexibility and scalability, providing effective support for distributed intelligent scheduling, cross-node resource coordination and management, multi-capability integration, network fault tolerance enhancement.
\vspace{-1.0em}
\subsubsection{Intelligent Fusion Decision}
\vspace{-0.5em}
In CIS-RAN, AI is introduced with a hybrid centralized-distributed deployement and management, enabling real-time and high-precision intelligent decision-making. 
Specifically, the AI node enables centralized management of the resource pools across multiple eBBUs to optimize load balancing and resource sharing globally, leveraging powerful computing resources for global model updates.
The distributed computing power at eBBUs support distributed intelligent scheduling and distributed training since the of real-time network status data can be interacted through flexible interfaces.
Such a hybrid centralized-distributed deployment and management of intelligent resources facilitates data-driven decision-making, dynamic resource optimization, and business intelligent adaptation.
It enables the shift from a "passive response" to a "proactive service" paradigm, driving the evolution of communication networks towards self-awareness, self-decision-making, and self-optimization.
\vspace{-1.0em}
\section{Key Technologies of CIS-RAN}\label{sec:technologies}
\vspace{-1.0em}
The essence of the CIS-RAN architecture lies in enabling cooperative communication, cooperative sensing, and cooperative AI.
At the core of these functions is a fundamental process comprising several key steps across the entire information lifecycle, including acquisition, transmission, and processing.
In the rest of this section, we illustrate and discuss representative key technologies corresponding to each of these steps.
\vspace{-2.5em}
\subsection{Efficient Information Acquisition}
\vspace{-0.5em}
With the emergence of multi-dimensional capabilities, achieving optimal information acquisition and transmission to meet the diversified demands of different usage scenarios, including communication and sensing, becomes a key challenge in the design of the future RAN architecture.
For example, sensing data differ significantly from communication data in terms of format, latency, and reliability requirements.
It is critical to realize low-overhead acquisition and transmission of different types of data while maintaining the accuracy, timeliness, and integrity of the information in CIS-RAN architecture design.
    
\begin{figure}[t]
    \centering
    \includegraphics[width=0.45\textwidth]{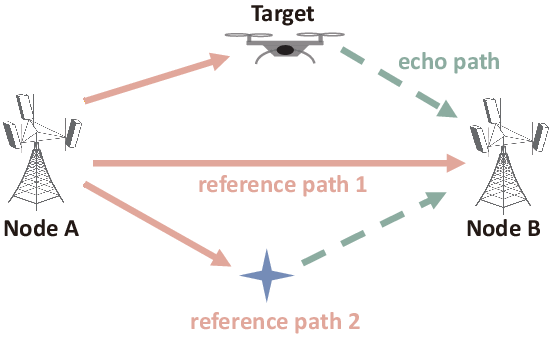}
    \caption{Schematic diagram of time-frequency synchronization.}
    \label{Fig:time-fre-syn}
\end{figure}

For sensing applications such as environment reconstruction and target detection, the most important thing is to obtain the distance and Doppler/velocity information of each path. In this regard, as shown in Figure~\ref{Fig:time-fre-syn}, we first obtain the received signals of all paths, estimate the distance and Doppler/velocity information of each path, and then calibrate the sensing results based on the synchronization method of the reference path, so as to obtain accurate sensing results.
With compensation of both timing and frequency synchronization obtained by using the line-of-sight~(LoS) path, the estimation performance of distance and velocity of the target can be much improved as shown in Table~\ref{tab:sensing}

To enhanced cooperative communication, it provides a novel method for channel state information CSI estimation. Compared to traditional methods, it offers significant advantages in accuracy and robustness. By leveraging the reference path for synchronization, both time and frequency errors are effectively compensated, leading to more precise distance and velocity estimation. This enhances the reliability of environmental reconstruction and target detection in cooperative sensing systems. Additionally, the proposed method reduces the computational complexity associated with CSI estimation, making it a more efficient and practical solution for 6G applications.
\begin{table}[t]
    \small
    \centering
    \caption{Numerical result of time-frequency synchronization}
    \begin{tabular}{c|m{1.5cm}<{\centering}|m{1.5cm}<{\centering}}
        \hline
        \hline
         & \textbf{Projected velocity} & \textbf{Prorogation distance}\\
        \hline
        \hline
        Actual value & $20$ m/s & $947.2$ m\\
        Estimated value with compensation & $26.77$ m/s & $1248$ m\\
        Estimated value w/o compensation & $18.46$ m/s & $946.7$ m\\
        \hline
        \hline
    \end{tabular}   \label{tab:sensing}
\end{table}
\vspace{-0.5em}
\subsection{Diverse Cooperation and Interaction}
To achieve broader and deeper network cooperation, higher demands are placed on the information interaction capabilities of future RAN architectures under latency, bandwidth, and topology constraints.
First, existing transmission protocols incur additional overhead when supporting real-time applications that integrate multi-dimensional capabilities, such as autonomous driving.
Second, considering limited bandwidth, high-resolution sensing data, such as 8K video, may cause a significant surge in the fronthaul interface.
Third, the topology of existing RAN architectures causes significant signaling overhead, as exemplified by the inability to address inter-BS L1/L2-triggered mobility~(LTM) due to the extremely high signaling overhead of frequent handovers, especially with higher mobility requirements in 6G.
    
The existing 5G RAN protocol design is oriented to communication purposes but less to the needs of AI/ML and sensing data transfer between user equipment~(UE) and RAN.
Several limitations in reusing the 5G signaling radio bearers~(SRBs) and data radio bearers~(DRBs) have been analyzed in 3rd generation partnership project~(3GPP) TR 38.843 as well as other researches~\cite{study2023technical}.
The SRBs have a higher priority in transmission, but the maximum payload size is restricted to $9000$ bytes per transaction, which cannot support large volumes of AI/ML or sensing data.
The DRBs have no restrictions in data size, but is subject to the configurations (e.g., QoS) and charging in the core network, making it less favorable for AI/ML and sensing tasks in RAN. Moreover, when a large volume of data is collected over the air, there exists a risk in draining the radio link and impeding the basic communication services in RAN. 

To address these issues, a novel type of radio bearers may be defined for the next generation RAN~\cite{AI2}. Such a type of radio bearers is solely managed by RAN, which allows RAN to establish and configure it on the demands of AI/ML and sensing tasks in RAN. Those radio bearers are capable of delivery of multiple payload types, including but not limited to, the sensing data, the AI/ML data such as the training data as well as the AI/ML models per se, or other enhancement data that may assist the optimization of AI/ML and sensing tasks. The design of such radio bearers involves a robust framework that can dynamically adapt to the varying characteristics of the delivered data, such as its type, volume, and priority. It is expected that the new radio bearers will ensure both flexibility and efficiency in AI/ML and sensing tasks. 

In addition, the scheme can be further extended to support the exchange of AI/ML and sensing data among the UEs or the RAN nodes, to enable more advanced AI/ML and sensing collaborations for RAN.    
\vspace{-0.5em}
\subsection{Intelligent Fusion Decision}
The enhanced network cooperation relies on intelligent decision-making capabilities, leading to a surge in computational power demands~\cite{han2020artificial}.
Optimizing model algorithms, as well as the deployment and management of computational resources, becomes crucial to enhance network accuracy and real-time performance.
In traditional resource management and deployment strategies, the centralized ones face challenges such as single-point malfunction, limited scalability, and high latency, especially as the network cooperation scale expands.
While distributed management can effectively address these issues, it faces problems like resource competition, redundancy, and poor data consistency.
Therefore, future RAN calls for a novel network architecture to realize intelligent fusion decision-making, which involves integrating information from various domains, including communication, sensing, and computing, to make informed decisions and optimize system performance.

One key aspect of intelligent fusion decision-making is the use of advanced machine learning algorithms. These algorithms can analyze the collected data, identify patterns, and extract valuable insights. By leveraging these insights, CIS-RAN can make decisions regarding resource allocation, beamforming, and other network parameters. This ensures that the network operates efficiently and effectively, meeting the diverse requirements of different applications and dynamic scenarios. For instance, machine learning can be employed to predict user mobility patterns, allowing for proactive resource allocation and seamless handover~\cite{lu2024deep}. Additionally, it can analyze sensing data to identify obstacles and optimize beamforming accordingly, enhancing communication quality and reliability.

Another crucial aspect of intelligent fusion decision-making is the ability to adapt and learn from evolving network conditions. CIS-RAN incorporates mechanisms for continuous learning and adjustment, allowing it to adapt to changing user demands, environmental conditions, and emerging technologies. This ensures that the network remains efficient and effective over time, even as the surrounding environment evolves. By combining advanced machine learning algorithms with adaptive learning mechanisms, CIS-RAN achieves intelligent fusion decision-making. This enables it to make informed decisions, optimize network performance, and deliver enhanced communication, sensing, and computing capabilities across diverse applications and dynamic scenarios.
\vspace{-1em}
\section{Case Study: CIS-RAN for MIMO}\label{sec:case}
\vspace{-1em}
Leveraging the key technologies discussed above, cooperative MIMO in CIS-RAN architecture is introduced, focusing on CSI acquisition, dynamic clustering, resource allocation, precoding design, and simulation results, respectively.

\begin{figure}[t]
	\centering
    \includegraphics[width=0.45\textwidth]{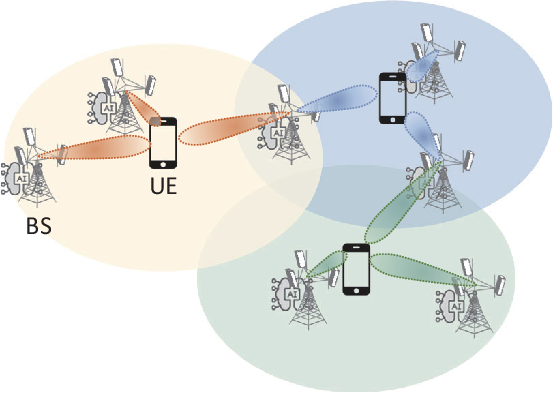}
	\caption{System model of cooperative MIMO in CIS-RAN.}
	\label{Fig:casemodel}
\vspace{-1.5em}
\end{figure}
In the rest of this section, as shown in Figure~\ref{Fig:casemodel}, we consider a downlink communication system consisting of $N$ BSs\endnote{Each eRRU $n$ - eBBU $m$ link is viewed as a BS. In CIS-RAN architecture, multiple BSs may share the same functional module, such as the MAC.}, each equipped with $N_t$, antennas, and $K$ single-antenna UEs.
To mitigate inter-cell interference, a user-centric MIMO architecture is employed where a set of candidate BSs is coordinated to provide wireless services to each UE~$k$, forming the $k$-th user-centric cluster.
Benefiting from the hierarchical aggregation and distributed intelligence, the fronthaul load and computational complexity can be reduced in CIS-RAN, while the classified acquisition and transmission provide a efficient way to acquire CSI.
\vspace{-0.5em}
\subsection{CSI Acquisition}
With the integration of sensing capability, network nodes in CIS-RAN are able to regularly transmit sensing signals like pilots or reference signals to model and reconstruct the surrounding environment~\cite{10872780}. Sequentially, abundant semantic characteristics of the channel are acquired.
Specifically, the environment objects (EOs) reconstructed by sensing are the multipath components (MPCs) that cause the scattering, reflection, and refraction of the signal propagation.
The parameters of each MPC containing the angle-of-arrival (AoA), angle-of-departure (AoD), delay, radar cross section (RCS) and received signal strength indicator (RSSI) can be estimated by sensing algorithms~\cite{10557715,10615717}.
The above CSI can then be used to assist communication.
By collecting the huge amount of sensing data and CSI, and with the help of AI processing, a wireless signal propagation map is able to be established.
The wireless map not only offers prior channel information to enable low-cost CSI acquisition but also fosters a shift from a "passive response" to a "proactive service" paradigm, accelerating the evolution of communication networks toward self-awareness, intelligent decision-making, and autonomous optimization.
 \vspace{-1.0em}
\subsection{Dynamic Clustering}
\begin{figure}[t]
	\centering
    \includegraphics[width=0.45\textwidth]{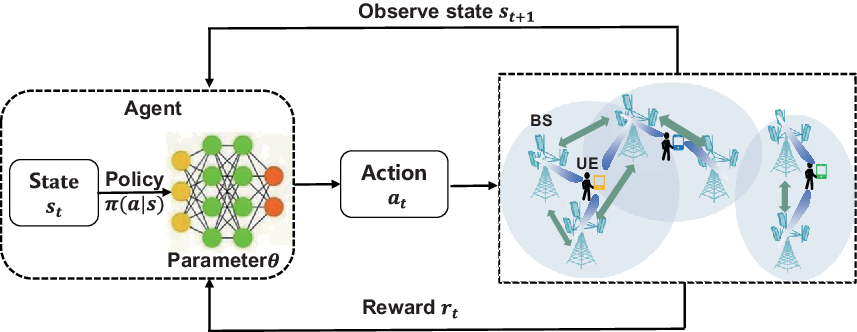}
	\caption{Dynamic clustering of MIMO in CIS-RAN.}
	\label{Fig:cluster}
    \vspace{-1.5em}
\end{figure}
To mitigate inter-cell interference, a straightforward approach is to coordinate all geographically distributed BSs to serve UEs which incurs an unacceptably high cost.
Therefore, in C-RAN, a predefined group of BSs is assigned to serve UEs within a constrained geographical area, forming static clusters that remain vulnerable to inter-cluster interference and fail to adapt to real-time variations in user demand, channel conditions, and network traffic.
To overcome this limitation, in CIS-RAN, each UE selects a subset of BSs based on criteria such as serving distance or network performance, breaking geographical static constraints and forming a dynamic cluster.

Benefiting from the hybrid centralized-distributed deployment of intelligent computing power in CIS-RAN, AI-enabled clustering becomes a promising technique to solve such a large-scale combinational optimization problem~\cite{biswas2021ap}.
Specifically, AI methods, particularly deep reinforcement learning (DRL), have emerged as a powerful tool for achieving intelligent and adaptive clustering~\cite{banerjee2023access}.
By leveraging DRL, dynamic clustering can be performed in a way that continuously learns from real-time network conditions, such as user locations, channel states, and traffic demands, to form optimal BS clusters and assign users to the most suitable BSs~\cite{gao2024drl}.

Each eBBU is viewed as an agent while the environment including real-time CSI, user location, quality of services, etc.
The connection between BSs and UEs is viewed as the action to be performed by the agent, which involves selecting the optimal subset of BSs from the candidate set for each user to form a dynamic cluster.
The reward function can include metrics such as spectral efficiency and energy efficiency.
Additionally, the potential interference caused by the current cluster to other clusters can be incorporated as a penalty term to avoid inter-cluster interference.
Each time the agent performs an action, it receives a corresponding reward, and the environment changes accordingly.
The agent observes and interacts with the environment, learning through this interaction.
In the next cycle, the agent selects the action that maximizes the reward, ultimately constructing a dynamic collaboration cluster centered around the user.
\vspace{-1.0em}
\subsection{Precoding Design}
\vspace{-0.5em}
With conventional RAN architectures, the precoding design of cooperative MIMO relies on the aggregation and processing of data from all UEs and the complete CSI between all BSs and UEs, leading to the heavy burden of fronthaul and extremely high computational complexity of precoding design, especially with as the network scales~\cite{zheng2024mobile}.
To tackle the challenge, CIS-RAN makes distributed precoding design possible benefiting from flexible interfaces, diverse interactions, and distributed intelligence~\cite{zhang2025a,shi2011iteratively}.

For each UE~$k$, the BS with the highest reference signal received power~(RSRP) is regarded as the master BS and responsible for the precoding design within current cluster.
To avoid extremely high computational complexity, we decouple the sum rate maximization problem into multiple parallel subproblems, which are solved independently by different clusters in a distributed manner~\cite{zhang2025a}.
However, due to the lack of global information, each cluster designs its own precoding matrices independently, resulting in inter-cluster interference.
To cope with this issue, the interference cost matrix~\cite{lagen2015distributed} is introduced to depict the equivalent interference channel towards UEs in neighboring clusters, which can be interacted via interfaces among eBBUs in CIS-RAN.
Each subproblem can be computed on distributed intelligent nodes, with the results uploaded to the AI node for global optimization.
With such a distributed precoding, network cooperative MIMO in CIS-RAN outperforms due to higher resource utilization rate and reduced computational complexity~\cite{park2012weighted,golub2013matrix}, as analyzed in our previous work~~\cite{zhang2025a}.
\vspace{-1.0em}
\subsection{Simulation Results}
In this subsection, we evaluate the throughput performance of cooperative MIMO in CIS-RAN.
For the propagations, we use the UMi path loss model certified by 3GPP~\cite{generation2019technical}.
The small-scale fading is modeled as Rician fading channel model. 
For comparison, the following algorithms are performed as well.
1) \emph{Single-node MIMO}: A conventional cellular network where each UE is served by a single BS.
2) \emph{Cooperative MIMO}: Under the boundary constraints of C-RAN, multiple geographically distributed BSs cooperate to mitigate inter-cell interference through a network-centric implementation.
3) \emph{Network cooperative MIMO (centralized)}: By breaking the boundaries of traditional C-RAN, a large number of geographically distributed BSs interconnected via fronthaul links to a CPU can cooperatively serve UEs across the entire coverage area through performing CJT.
4) \emph{Network cooperative MIMO (distributed)}: To address this issue, the network-wide optimization problem is decoupled into multiple parallel cluster-level subproblems~[42]. Each subproblem is independently solved by a cluster in an iterative manner until convergence.
\begin{figure}[t]
	\centering
    \includegraphics[width=0.45\textwidth]{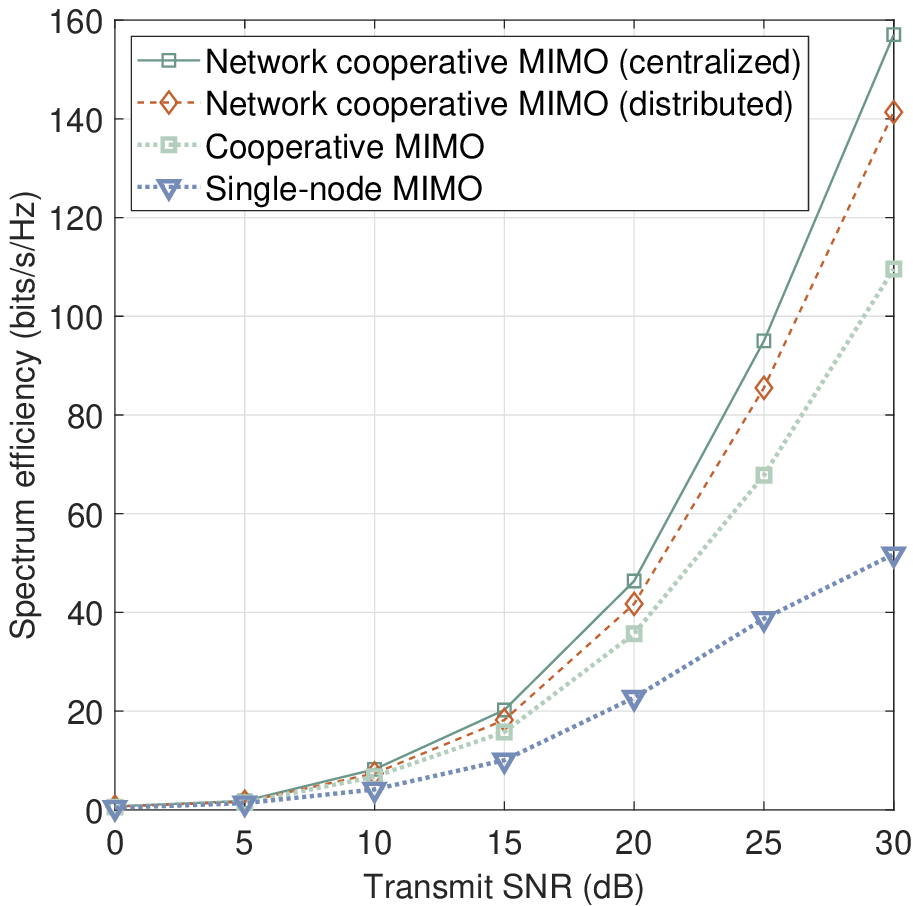}
	\caption{Spectrum efficiency vs. transmit SNR by adopting different MIMO techniques.}
	\label{Fig:rate}
    \vspace{-1.5em}
\end{figure}
\begin{figure}[t]
	\centering
    \includegraphics[width=0.45\textwidth]{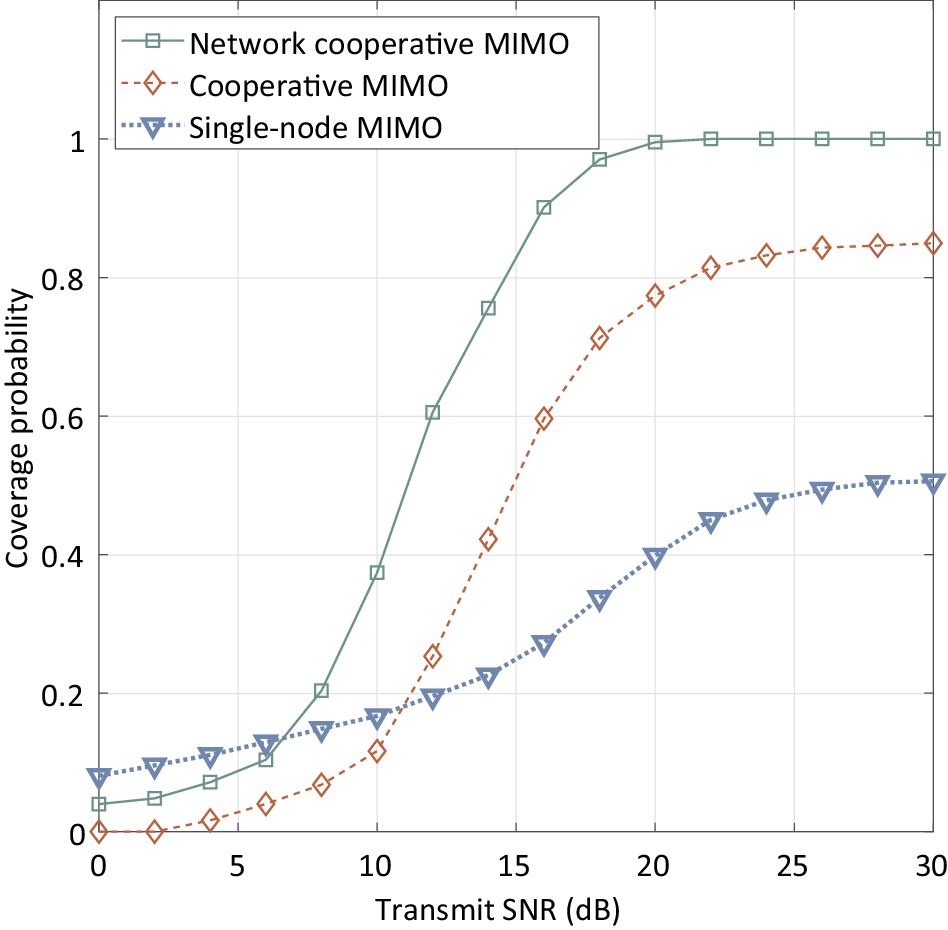}
	\caption{Coverage probability vs. transmit SNR by adopting different MIMO techniques.}
	\label{Fig:coverage}
\vspace{-1.5em}
\end{figure}
\begin{figure}[t]
	\centering
    \includegraphics[width=0.45\textwidth]{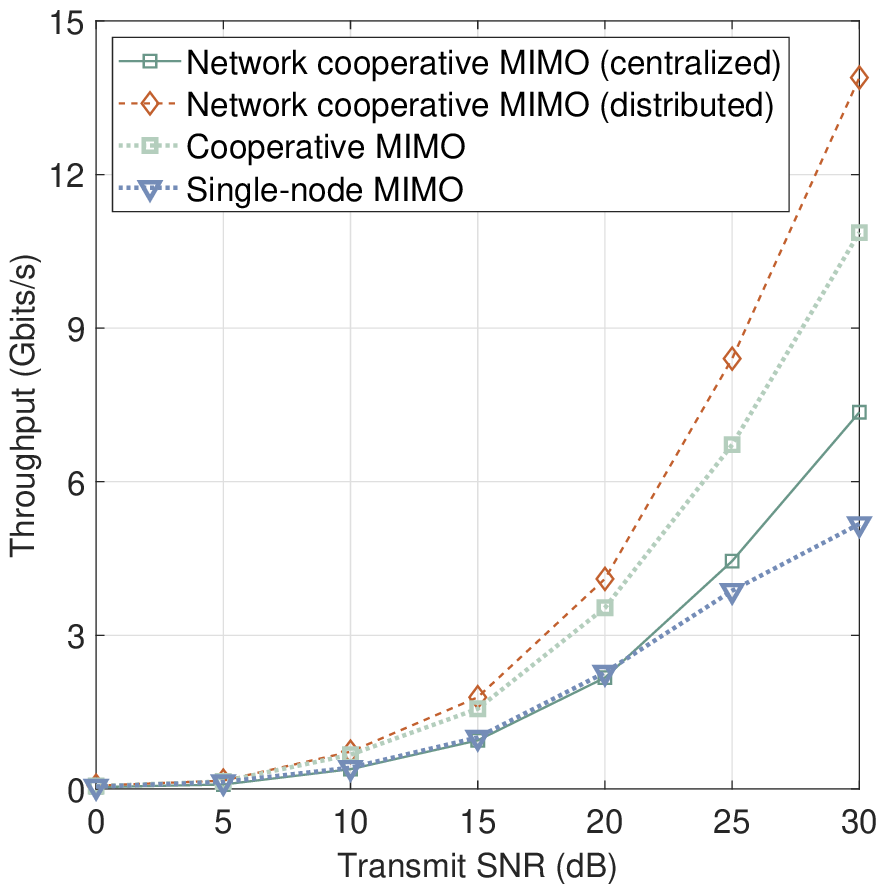}
	\caption{Throughput vs. transmit SNR by adopting different MIMO techniques.}
	\label{Fig:tp}
\vspace{-1.5em}
\end{figure}
\begin{figure}[t]
	\centering
    \includegraphics[width=0.45\textwidth]{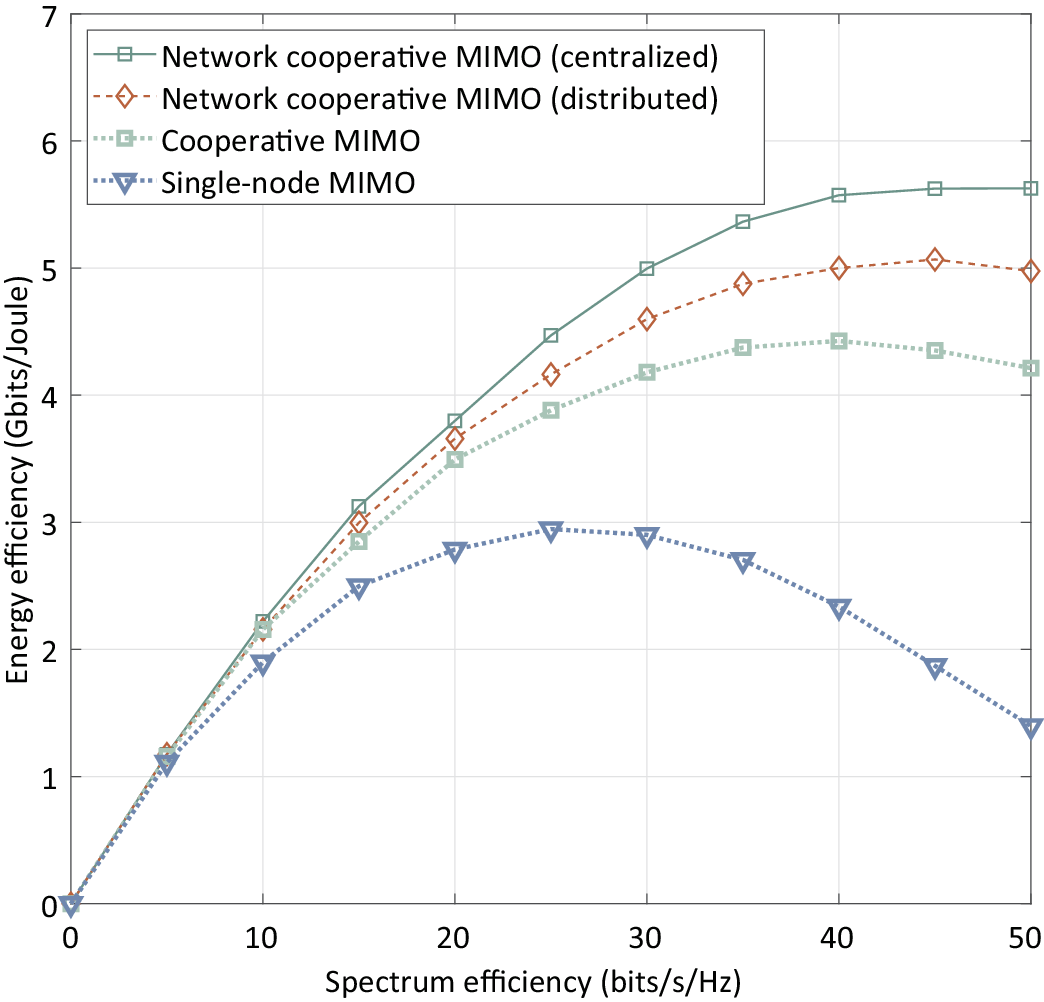}
	\caption{Energy efficiency vs. spectrum efficiency by adopting different MIMO techniques.}
	\label{Fig:EE}
\end{figure}
Figure~\ref{Fig:rate} shows the spectrum efficiency versus the transmit SNR with $32$ UEs.
It can be observed that the spectrum efficiency of different MIMO techniques increases with the transmit SNR.
Enabled by the CIS-RAN architecture, network cooperative MIMO achieves significantly higher spectrum efficiency than conventional ones by effectively mitigating interference through user-centric coordination.
Furthermore, the distributed algorithm benefits from the flexible interfaces, rich interactions, and distributed intelligence supported by CIS-RAN, achieving performance close to the centralized algorithm while incurring much lower computational complexity, as analyzed in our previous work~\cite{zhang2025a}.

Figure~\ref{Fig:coverage} illustrates the coverage probability achieved by various MIMO techniques under different transmit SNR, where coverage refers to the received SINR exceeds a given threshold~\cite{bai2014coverage}.
It can be seen that network cooperative MIMO achieves substantially better coverage performance than conventional techniques by suppressing interference and ensuring a consistent user experience through user-centric coordination.
With a low transmit SNR, single-node MIMO exhibits slightly higher coverage probability than the two cooperative MIMO techniques, as the BS can concentrate its limited power on nearby users.
In contrast, the power-sharing nature in cooperative techniques may lead to reduced cooperation gain under power constraints.
As the transmit SNR increases, the cooperation gain becomes more pronounced, resulting in a rapid improvement in coverage probability for both cooperative MIMO techniques.

Given $t$ time slots including the time for precoding design and that for data transmission, the throughput is calculated by~$T = \frac{t-\frac{\mathcal{O}}{c}}{t}BR$, where $B$ represents the bandwidth.
The notation~$\mathcal{O}$ and $c$ denote the computational complexity of precoding design and the computing capacity which is dependent by the maximum CPU cycles per second, respectively.
Figure~\ref{Fig:tp} illustrates the throughput performance versus transmit SNR for different MIMO techniques, where $16$ BSs serve $32$ UEs over a $100$~MHz bandwidth.
By jointly considering both data rate and computational complexity, it can be observed that network cooperative MIMO achieves higher throughput under a distributed algorithm, as it enables parallel processing across clusters while effectively mitigating interference.
In contrast, with a centralized algorithm, the throughput is significantly constrained due to the high computational complexity of joint processing at the CPU, particularly in large-scale deployments.

Figure~\ref{Fig:EE} illustrates the energy efficiency versus spectrum efficiency for different MIMO techniques over a $100$~MHz bandwidth.
The power consumption consists of the static hardware power of each BS, which remains constant, and the transmit power, which varies with the adopted precoding and power allocation schemes.
It can be observed that as the spectrum efficiency increases, the energy efficiency first rises and then declines. Compared to conventional MIMO techniques, network cooperative MIMO achieves higher energy efficiency across a range of spectrum efficiency levels.
\vspace{-1.0em}
\section{Future Works and Standardization Issues}\label{sec:future}
\vspace{-0.5em}
By leveraging hierarchical aggregation, classified data acquisition and transmission, and distributed intelligence, CIS-RAN achieves broader and deeper network cooperation, enabling more efficient information acquisition, diverse cooperation and interaction, and intelligent fusion decision, thus offering valuable insights for the future of network development and standardization, which are discussed as below.
\vspace{-1.0em}
\subsection{Candidate RAN Functional Split Options}
The fundamental C‑RAN advanced architecture is developed in Section~\ref{sec:fundamentals}, yet several key design details remain to be addressed.
A primary consideration is the decomposition of the eBBU into an enhanced CU~(eCU) and an enhanced DU~(eDU).
As shown by the green dashed box in Figure \ref{Fig:CRANa}, such a split would allocate centralized control and high‑level protocol processing to the enhanced CU, while the enhanced DU assumes distributed processing of lower‑layer protocols.
The architecture reorganization promises to improve network flexibility and scalability, while enabling intelligent distributed scheduling, capability integration, stronger fault tolerance, and cost efficiency.  
With the adoption of the eCU–eDU–eRRU architecture, it becomes crucial to explore how to optimally deploy sensing and distributed AI functions, including decisions on component deployment and standardized evolution path.
\vspace{-1.0em}
\subsection{Interaction Mechanism Design}
Building on the CIS-RAN architecture, interaction mechanisms across the data, user, and computing planes must be carefully designed to support efficient network cooperation.
Currently, UEs conduct SSB measurements and complete the random access procedure with a single serving BS.
Cooperative transmission is only supported during the data transmission phase in the RRC-connected state, limiting the ability to achieve instant access or seamless reconfiguration without service interruption.
CIS-RAN breaks through the physical and logical constraints of traditional C-RAN, enabling new possibilities for low-latency interactions.
To fully exploit these capabilities, it is essential to design efficient, latency-aware interaction mechanisms.
Meanwhile, features such as always-on access and multi-point cooperative transmission challenge conventional UE access procedures, requiring updates to synchronization signal block~(SSB) design, RRC signaling, and the dynamic formation of cooperative clusters.
\vspace{-1.0em}
\subsection{Heterogeneous Hardware Implementation}
While CIS-RAN enhances cooperative communication, it also integrates cooperative sensing and AI, which necessitates compatibility with heterogeneous hardware components such as various processors, accelerators, and sensors, thereby increasing the complexity of hardware integration.
It is essential to ensure interoperability and efficient collaboration among these components, thereby enabling high-performance network cooperation.
Furthermore, to accommodate the diverse and dynamic resource demands of various network tasks, efficient hardware resource management mechanisms covering allocation, scheduling, and load balancing must be developed to improve utilization and reduce operational costs.
\vspace{-1.0em}
\subsection{Security and Privacy Issues}
CIS-RAN significantly enhances network performance through classified data acquisition, hierarchical aggregation, and distributed intelligence.
However, the enhancement of network cooperation also raises potential security challenges.
\emph{First}, the integration of sensing capabilities may collect sensitive data like user location, behavior, and biometric information, which could lead to privacy violations~\cite{shao2008pdcs}.
Lightweight encryption algorithms, such as AES-GCM~\cite{mozaffari2011efficient}, and federated learning can help mitigate these concerns by protecting sensitive data and preserving privacy.
\emph{Second}, the introduction of diverse interfaces increases the risk of man-in-the-middle attacks, data tampering, and DoS threats.
\emph{Third}, AI-based intrusion detection systems~(IDS)~\cite{rajapaksha2023ai} and defensive strategies such as interface authentication and secure transmission can address these risks.
Moreover, adversarial attacks on AI models, such as falsified traffic features, can lead to biased results.
Injecting adversarial samples during training and using trusted execution environments~(TEEs) can enhance robustness and prevent tampering~\cite{jauernig2020trusted}.
Other challenges, such as balancing privacy protection with service performance and ensuring real-time security, also require further attention.
\vspace{-0.5em}
\section{Conclusion}\label{sec:conclusion}
To enhance performance and effectiveness across diverse application scenarios, in this article, we proposed the CIS-RAN architecture, designed to expand network cooperation in terms of both broader geographical coverage and deeper fusion of multi-dimensional capabilities.
By incorporating classified data acquisition, hierarchical aggregation, and distributed intelligence, CIS-RAN facilitates efficient information acquisition, diverse data interactions, and intelligent fusion decision-making. 
Key technologies were discussed, with network cooperative MIMO used as a case study, showcasing significant performance gains compared to traditional architectures, as evaluated by numerical simulations. 
Future works were also outlined, focusing on further advancing the CIS-RAN architecture.
\vspace{-0.5em}
\section*{ACKNOWLEDGEMENT}
\label{ACKNOWLEDGEMENT}

*****************************

\theendnotes
\vspace{-0.5em}
\bibliographystyle{gbt7714-numerical}
\bibliography{myref}





\end{document}